%% file: main.tex
\documentclass[conference]{IEEEtran}
\IEEEoverridecommandlockouts

\input{preamble}

\begin{document}

\placetextbox{0.5}{0.99}{\large\colorbox{gray!3}{\textcolor{WildStrawberry}{\textbf{Author pre-print.}}}}%

\placetextbox{0.5}{0.97}{\large\colorbox{gray!3}{\textcolor{WildStrawberry}{Publication accepted for (\hreff{https://conf.researchr.org/home/models-2025}{MODELS'25}).}}}%

\placetextbox{0.5}{0.05}{\colorbox{gray!3}{\textcolor{WildStrawberry}{Author pre-print. Publication accepted for} \hreff{https://conf.researchr.org/home/models-2025}{MODELS'25}.}}%

\title{Complex Model Transformations by Reinforcement Learning with Uncertain Human Guidance}

\author{
     \IEEEauthorblockN{Kyanna Dagenais\IEEEauthorrefmark{1}\orcidlink{0009-0007-6304-3971},
     Istvan David\IEEEauthorrefmark{1}\IEEEauthorrefmark{2}\orcidlink{0000-0002-4870-8433}}
     \IEEEauthorblockA{\IEEEauthorrefmark{1}McMaster University, Hamilton, Canada}
     \IEEEauthorblockA{\IEEEauthorrefmark{2}McMaster Centre for Software Certification, Hamilton, Canada}
}

\maketitle

\input{sections/abstract}
\input{sections/intro}
\input{sections/background}
\input{sections/example}
\input{sections/approach}
\input{sections/guidance}
\input{sections/evaluation}
\input{sections/discussion}
\input{sections/conclusion}
\input{sections/acknowledgement}

\printbibliography

\end{document}

%% file: preamble.tex
\usepackage[utf8]{inputenc}
\usepackage{hyperref}
\usepackage[dvipsnames]{xcolor}
\usepackage{fdsymbol}
\usepackage{enumitem}
\usepackage[backend=biber,style=ieee,maxcitenames=2,maxbibnames=99,citestyle=numeric-comp]{biblatex}
\usepackage{orcidlink}
\usepackage{booktabs}
\usepackage{listings}
\usepackage{float}
\usepackage{multicol}
\usepackage{multirow}
\usepackage{makecell}
\usepackage{subcaption}
\usepackage[framemethod=TikZ]{mdframed}
\usepackage{nicefrac}

\newcommand{\secref}[1]{Sec.~\ref{#1}}
\newcommand{\figref}[1]{Fig.~\ref{#1}}
\newcommand{\tabref}[1]{Tab.~\ref{#1}}
\newcommand{\equref}[1]{Equation~\ref{#1}}
\newcommand{\lstref}[1]{Listing~\ref{#1}}

\newcommand{\viatra}{\textsc{Viatra}}

\newtheorem{definition}{Definition}%

\newlength{\dpcircle}
\newlength{\rcircle}
\newlength{\dcircle}
\newlength{\scaledcircle}
\newcommand{\docircle}[4]{%
  \setlength{\dpcircle}{\dp\strutbox}%
  \setlength{\dcircle}{\dpcircle}%
  \addtolength{\dcircle}{\ht\strutbox}%
  \setlength{\scaledcircle}{1.25\dcircle}%
  \setlength{\rcircle}{0.5\scaledcircle}%
  \setlength{\unitlength}{1sp}%
  \begin{picture}(\number\scaledcircle,0)
    \color{#1}
    \put(\number\rcircle,\number\dpcircle){\circle*{\number\scaledcircle}}
    \color{#2}
    \put(\number\rcircle,\number\dpcircle){\circle{\number\scaledcircle}}
    \put(\number\rcircle,0){\makebox[0pt]{\textcolor{#3}{#4}}}
  \end{picture}%
}

\newcommand{\colornumberdomainexpert}[1]{%
  {\footnotesize\docircle{domaingreen}{black}{black}{\textbf{#1}}}%
}

\newcommand{\colornumberauto}[1]{%
  {\footnotesize\docircle{autoblue}{black}{black}{\textbf{#1}}}%
}

\newcommand{\coord}[1]{%
  [#1]%
}

\newcommand{\cell}[2]{%
    \texttt{[{#1},{#2}]}%
}

\newcommand{\parameter}[3]{%
    \texttt{\cell{#1}{#2}, {#3}}%
}

\newcommand\bcf[1]{{#1}^{\odot}}

\newenvironment{conclusionframe}[1]
  {\mdfsetup{
    innertopmargin=4pt,
    frametitleaboveskip=-\ht\strutbox,
    frametitlealignment=\center
    }
  \begin{mdframed}%
  }
  {\end{mdframed}}

\hypersetup{pdfborder={0 0 0}}

\mdfsetup{%
   backgroundcolor=gray!10,
   middlelinewidth=1pt,
   roundcorner=10pt}

\definecolor{autoblue}{HTML}{cce4ff}
\definecolor{darkblue}{rgb}{0.0,0.0,0.6}
\definecolor{darkred}{rgb}{0.6,0.0,0.0}
\definecolor{domaingreen}{RGB}{209, 233, 143}
\definecolor{eclipsekeywordred}{RGB}{127,0,85}
\definecolor{mauve}{rgb}{0.58,0,0.82}
\definecolor{xtendorange}{RGB}{171,48,0}

\lstdefinelanguage{gridworld-dsl}
{
    comment=[l]{--},
    morekeywords={READ},
    keywordstyle=\color{white},
    morestring=[s][\color{darkred}\bfseries]{\{}{\}},
    morestring=[s][\color{darkred}\bfseries]{`}{`},
    morestring=[s][\color{mauve}\bfseries]{<}{>},
    extendedchars=true
}

\lstdefinelanguage{VQL}
{
    keywordstyle=[1]{\color{eclipsekeywordred}\bfseries},
    keywordstyle=[2]{\color{darkblue}\bfseries},
    morekeywords=[1]{pattern, find},
    morekeywords=[2]{Agent, Environment, TerminatingState},
    alsoletter={.},
    morekeywords=[2]{Environment.agent},
    extendedchars=true
}

\lstdefinelanguage{Xtend}
{
    keywordstyle=[1]{\color{eclipsekeywordred}\bfseries},
    keywordstyle=[2]{\color{darkblue}},
    keywordstyle=[3]{\color{xtendorange}\textit},
    morekeywords=[1]{val, it, override, return},
    morekeywords=[2]{agentIsTerminatedRule, AgentIsTerminatedRule, moveRightRule, MoveRightRule, nextActivation},
    morekeywords=[3]{createRule, agentInTerminatingState, agentInNonTerminatingState},
    morekeywords=[2]{container},
    extendedchars=true,
    literate={container}{{{\textcolor{darkblue}{container}}}}1,
    morekeywords=[2]{policyService},
    extendedchars=true,
    literate={policyService}{{{\textcolor{darkblue}{policyService}}}}2
}

\input{preamble/preprint}

%% file: preamble/preprint.tex
\usepackage{eso-pic}
\definecolor{linkblue}{HTML}{006fbd}
\newcommand{\hreffinternal}[3]{\href{#1}{\textcolor{#3}{#2}}}
\newcommand{\hreff}[2]{\hreffinternal{#1}{#2}{linkblue}}

\newcommand{\placetextbox}[3]{
  \setbox0=\hbox{#3}
  \AddToShipoutPictureFG*{
    \put(\LenToUnit{#1\paperwidth},\LenToUnit{#2\paperheight}){\vtop{{\null}\makebox[0pt][c]{#3}}}%
  }%
}%

%% file: sections/abstract.tex
\begin{abstract}
Model-driven engineering problems often require complex model transformations (MTs), i.e., MTs that are chained in extensive sequences.
Pertinent examples of such problems include model synchronization, automated model repair, and design space exploration. Manually developing complex MTs is an error-prone and often infeasible process. Reinforcement learning (RL) is an apt way to alleviate these issues. In RL, an autonomous agent explores the state space through trial and error to identify beneficial sequences of actions, such as MTs. However, RL methods exhibit performance issues in complex problems. In these situations, human guidance can be of high utility. In this paper, we present an approach and technical framework for developing complex MT sequences through RL, guided by potentially uncertain human advice. Our framework allows user-defined MTs to be mapped onto RL primitives, and executes them as RL programs to find optimal MT sequences. Our evaluation shows that human guidance, even if uncertain, substantially improves RL performance, and results in more efficient development of complex MTs. Through a trade-off between the certainty and timeliness of human advice, our method takes a step towards RL-driven human-in-the-loop engineering methods.
\end{abstract}


\newcommand{\badge}[3]{
	\ifthenelse{\equal{#1}{}}{}{
		\begin{tikzpicture}[overlay, remember picture]
			\node[xshift=-4.4cm,yshift=-1.1cm] at (current page.north east) {\includegraphics[width=1.5cm]{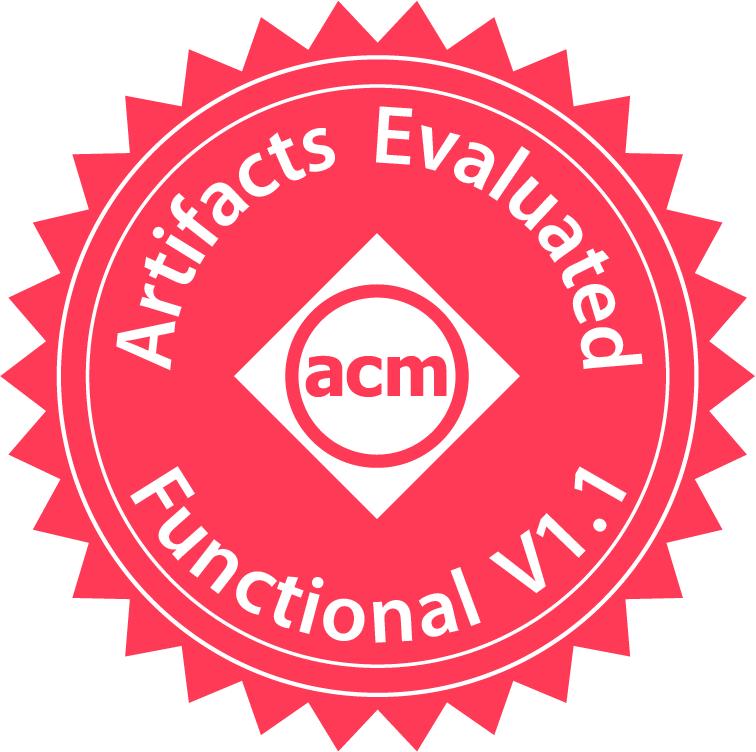}};
		\end{tikzpicture}}
	\ifthenelse{\equal{#2}{}}{}{
		\begin{tikzpicture}[overlay, remember picture]
			\node[xshift=-2.8cm,yshift=-1.1cm] at (current page.north east) {\includegraphics[width=1.5cm]{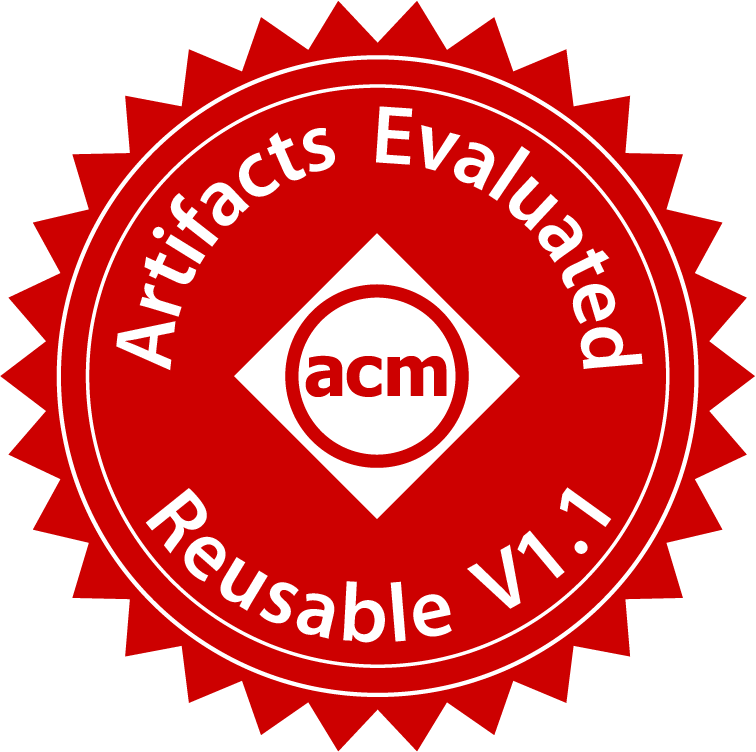}};
		\end{tikzpicture}}
	\ifthenelse{\equal{#3}{}}{}{
		\begin{tikzpicture}[overlay, remember picture]
			\node[xshift=-1.2cm,yshift=-1.1cm] at (current page.north east) {\includegraphics[width=1.5cm]{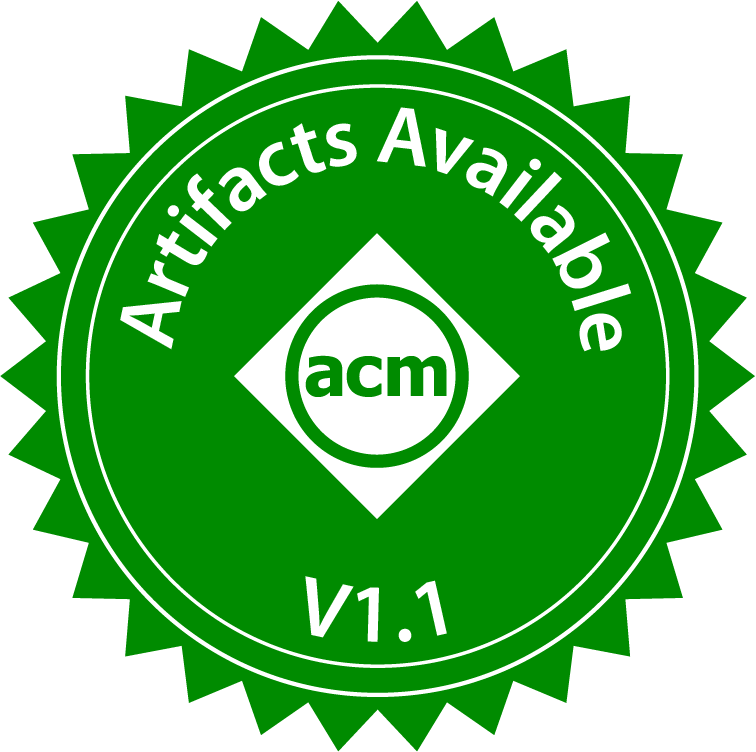}};
		\end{tikzpicture}}
}
\badge{}{reusable}{available}

%% file: sections/intro.tex
\section{Introduction} \label{sec:introduction}

Modeling activities are often more complex than an atomic model transformation (MT) and rely on \textit{sequences of MTs}. Pertinent examples can be found in model synchronization~\cite{marussy2018incremental}, model refactoring~\cite{mokaddem2018recommending}, and rule-based design-space exploration~\cite{abdeen2014multi}. Typically, there might be more than one MT sequence that can successfully transform the source model into the target state, and choosing the most appropriate (cost-effective, efficient, safe) one manually is not tractable. This raises the \textbf{need for automated methods for developing complex MTs}, in which MTs are chained in sequences.

Reinforcement learning (RL) is a machine learning (ML) paradigm, in which an agent explores the state space, and through trial and error, learns beneficial action sequences~\cite{sutton2018reinforcement}. This ability positions RL as a potential automation method for developing complex MTs.
The benefits of RL have been recognized in MDE before, e.g., in model repair~\cite{barriga2019personalized}, in-place MTs~\cite{eisenberg2021towards}, and model-driven service migration~\cite{dehghani2022facilitating}. However, RL methods perform poorly in the early phases of learning due to suboptimal decisions before converging to better policies~\cite{sutton2018reinforcement}. That is, RL agents may require substantial training time and resources before learning an optimal sequence of MTs. This shallow learning curve limits the applicability of RL, particularly in cases where early mistakes can be costly, e.g., synchronizing models in model-based design of safety-critical systems. While the performance of RL agents improves over time, this improvement often comes too late to be useful.

Human guidance is an apt technique to mitigate problems, such as the shallow learning curve of RL agents~\cite{najar2021reinforcement}. In such a setup, human advice acts as a heuristic to hint the otherwise autonomous agent towards desirable goals, thereby improving its performance. Involving human guidance in automated MDE methods has been explored before, e.g., driving RL-based MTs for model repair by personalized preferences~\cite{barriga2022parmorel}, and driving search-based methods for MT-based design space exploration~\cite{hegedus2015model}. A common theme of existing techniques is the assumption that firm advice can be produced by the human in a timely and cost-effective fashion. This is a strong assumption that cannot always be guaranteed, especially when the human is asked to provide guidance about complex MTs (e.g., identifying consistent intermediate states in a long refactoring, or known locally optimal configurations in design-space exploration).  Human advice, though often uncertain, can still be of high value in driving RL agents~\cite{dagenais2024towards}. This raises the \textbf{need for incorporating human advice into RL-driven MTs through sound modeling and quantification of uncertainty}.

In this paper, \textbf{we develop an approach and technical framework to infer complex MTs through RL, guided by potentially uncertain human advice}. The sound formalization and assessment of uncertainty allows for producing useful human advice early in the MT learning process. Depending on the specific situation and engineering problem, early uncertain advice might be of higher value than late certain advice. We evaluate our approach in synthetic and real human guidance scenarios. Our results indicate that human guidance, even if uncertain, substantially improves RL performance, and allows for the efficient inference of complex MTs.\footnote{A replication package with detailed replication instructions, results, and analyses is publicly available at \url{https://doi.org/10.5281/zenodo.16321499}.}

Through a trade-off between the certainty and timeliness of guidance, our method takes a firm step towards ML-driven human-in-the-loop engineering methods. Our approach alleviates the manual labor required for developing complex MTs while enabling domain experts to steer the inference process. This convergence of human and artificial intelligence is much sought-after in complex engineering domains.

%% file: sections/background.tex
\section{Background and Related work}\label{sec:background}

Here, we introduce the key concepts of our approach: reinforcement learning (\secref{sec:rl}) and its guidance by humans (\secref{sec:background-guidance}); and we discuss the related work (\secref{sec:relwork}).

\subsection{Reinforcement learning}\label{sec:rl}

In RL,
an \textit{agent} learns optimal control of an \textit{environment} by taking sequential actions to explore the environment and update its strategy based on feedback in the form of \textit{rewards}~\cite{sutton2018reinforcement}. This can be formalized as a Markov decision process~\cite{puterman1990markov} $\langle S, A, \pi(a|s), r_t(s, a)\rangle$. $S$ denotes the set of observable states the environment can produce, and $A$ is the set of actions the agent can take.  $r_t(s,a) \in \mathbb{R}$ is the reward the environment produces when the agent chooses action $a\in A$ while observing state $s\in S$, at time step $t$. At time step $t$, the agent observes a state $s_t$ and chooses an action $a_t$. In response, at time step $t+1$, the environment produces a reward $r_{t+1}$ and a new state $s_{t+1}$. The agent maintains a policy $\pi(a|s)$, which gives the conditional probability of choosing action $a$ in state $s$. The agent's goal is to learn and act according to an optimal policy $\pi_*$ that maximizes the sum of future rewards. 

RL has seen growing use in support of MDE, e.g., in MT engineering~\cite{eisenberg2021towards}, model repair~\cite{barriga2022parmorel}, fault injection testing~\cite{moradi2020exploring}, and inference of simulators for digital twins~\cite{david2022devs}. \textbf{In this work}, we use RL to learn complex MTs, i.e., long sequences of MTs that are infeasible to reason about for human modelers.

\subsection{Human guidance in RL}\label{sec:background-guidance}

While RL algorithms can operate autonomously, evidence suggests that human guidance is beneficial to learning dynamics. Guidance is a method in which human input is used to bias the agent's exploration strategy. Techniques differ by when and how advice is provided, as described in the taxonomy of \textcite{najar2021reinforcement}. For example, aligned with our approach, \textcite{cruz2017agent} emphasize the role of \textit{early advice}.

Strategies to incorporate guidance into RL are called \textit{policy shaping} methods, which are defined by which part of the RL process they are incorporated in. In value shaping, human advice acts as a function of action preferences~\cite{najar2016training}. In reward shaping, advice is translated into numerical rewards given to the agent~\cite{tenoriogonzalez2010dynamic}. In decision biasing, human advice is used to directly affect the policy output~\cite{thomaz2006reinforcement}. Finally, in policy shaping, human advice is directly incorporated into the agent's policy to bias the exploration strategy. For example, \textcite{griffith2013policy} estimate the human's feedback policy and incorporate this information into the agent's policy.

By this taxonomy, \textbf{our work} falls under the category of guidance, and specifically, \textbf{policy shaping}, since we aim to bias the agent's exploration by human guidance.

\subsection{Uncertainty and Subjective Logic}\label{sec:subjectiveLogic}
Using human guidance to bias the RL agent's policy requires a formal approach to appropriately capture human opinions. In particular, the epistemic uncertainty of human opinions must be modeled appropriately to make guidance meaningful. Subjective Logic (SL) \cite{j_osang2016subjective} is a formalism that defines an opinion, a construct composed of belief and an orthogonal component of uncertainty. A binomial opinion is defined as  $\omega_x =(b_x, d_x, u_x, a_x)$. This opinion is about the truth of a Boolean predicate $x$. Here $b_x$ is belief in $x$, $d_x$ disbelief in $x$, $u_x$ vacuity of evidence of $x$, and $a_x$ prior probability of $x$. The parameters must adhere to the constraints $b_x, d_x, u_x, a_x \in [0,1]$ and $b_x + d_x + u_x = 1$. A binomial opinion is transformed into a probability $P(x) = b_x +a_xu_x$. Likewise, probability $p(x)$ translates to binomial opinion $\omega = (p, 1-p, 0, p)$. SL defines fusion operators that allow for multiple opinions to be fused into one opinion.

The utility of SL in MDE has been demonstrated, e.g., in modeling with uncertain types~\cite{burgueno2018expressing, bagheri2009belief, troya2021uncertainty} and prioritizing inconsistency resolution steps~\cite{jongeling2023uncertainty}. \textbf{In this work}, we use SL to model the uncertainty of guidance for RL in MT development.

\subsection{Related Work}\label{sec:relwork}

Closest to our work are approaches that use RL for MTs, particularly with human in the loop for guidance or intervention. The PARMOREL framework \cite{barriga2022parmorel,iovino2020model} uses RL to learn user-preferred sequences of model transformations to fix invalid models.
While this approach features human input for choosing an appropriate reward function, it does not accommodate the uncertainty of human guidance.

\citeauthor{eisenberg2021towards}~\cite{eisenberg2021towards,eisenberg2024single} use RL for learning in-place MTs. In this approach, the RL agent learns a sequence of MTs that optimizes objectives in the final model. The authors demonstrate that RL is a feasible alternative to traditional search heuristics, e.g., genetic algorithms. Our work improves over these approaches by including human guidance in RL.

\textcite{basciani2018tool} introduce the CITRIC tool to recommend developers optimal MT sequences between a source and target metamodel. The approach ranks MT sequences through shortest path algorithms by two relevant criteria: metamodel coverage and information loss. Our approach is similar in aims but differs in technological solution, including the role of the human, who is an active advisor in our approach.

Finally, some adjacent techniques that automate MT development, but not by RL, include the following.
\textcite{eisenberg2024multi} use multi-objective optimization to explore MT sequences using meta-heuristic search algorithms. \textcite{abdeen2014multi} use a genetic algorithm for multi-objective optimization in rule-based design-space exploration. \textcite{mokaddem2018recommending} learn refactoring rules from examples using genetic programming.

%% file: sections/example.tex
\section{Illustrative example}\label{sec:example}

\begin{figure}[tbh]
    \centering
    \begin{subfigure}{.52\linewidth}
        \centering
        \includegraphics[trim={7cm 6cm 6cm 6.5cm},clip,width=\linewidth]{figures/lake-class-diagram.pdf}
        \caption{Metamodel}
        \label{fig:frozenlake-metamodel}
    \end{subfigure}%
    \begin{subfigure}{.47\linewidth}
        \centering
        \includegraphics[width=\linewidth]{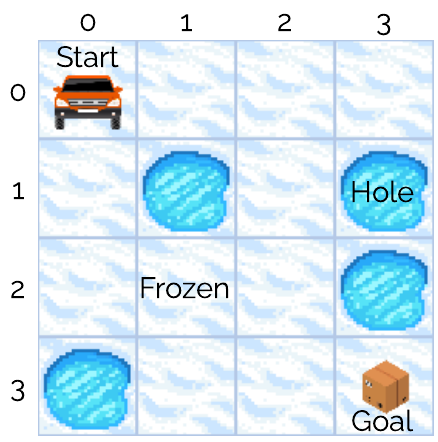}
        \caption{Example model}
        \label{fig:frozenlake-instance}
    \end{subfigure}
    \caption{The Frozen Lake running example}
    \label{fig:frozenlake}
\end{figure}

To illustrate our approach, we rely on the following running example of a self-driving vehicle through a frozen lake (\figref{fig:frozenlake}). The example is analogous to Open AI’s Gym's Frozen Lake environment\footnote{\url{https://gymnasium.farama.org/environments/toy_text/frozen_lake/}}, but it is fully modeled in EMF and its operational semantics are defined by MTs, akin to the Pacman case, a frequently used example in MDE~\cite{heckel2006graph,syriani2019domain-specific,eisenberg2021towards}.

The vehicle explores the lake by choosing to move up, down, left, or right one tile at a time. It begins its exploration at the top leftmost tile (\textit{Start}). Its aim is to reach the bottom rightmost tile (\textit{Goal}) while avoiding terminating states (\textit{Hole}s).

\subsection*{Model transformations}

\begin{figure}[tbh]
    \centering
    \includegraphics[trim={0.75cm 0.2cm 0.75cm 0.1cm},clip,width=\linewidth]{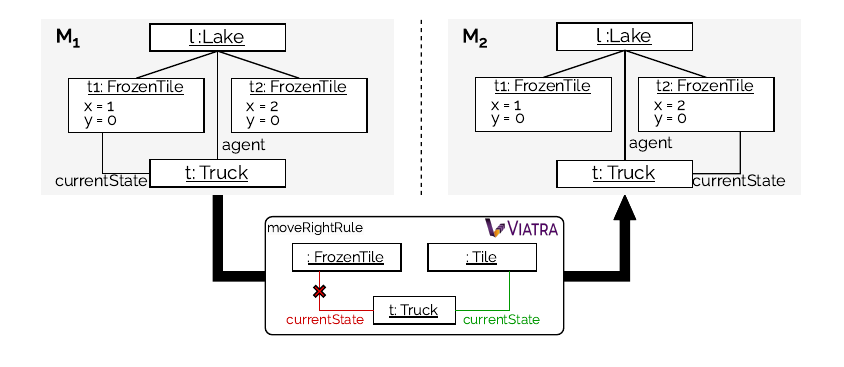}
    \caption{Example MT rule and its execution}
    \label{fig:mt-example}
\end{figure}

Operational semantics are given by MTs. As shown in \figref{fig:mt-example}, moving to the tile to the right is achieved by executing the \texttt{moveRightRule} MT rule, detailed in \lstref{lst:trafo-example}.
To illustrate our approach, we use the \viatra{} framework~\cite{bergmann2015viatra}.

\input{src/transformation-example}

To successfully navigate through the lake, the vehicle must find a MT sequence that brings it to the goal state while avoiding terminating states (holes).

\subsection*{Reinforcement learning}
In RL terms, the MT sequence will generate \textit{reward} upon encountering different states as follows.

\begin{description}
    \item[Frozen tile] Nothing happens. $\text{Reward}=0$.
    \item[Hole] The episode ends unsuccessfully. $\text{Reward}=0$.
    \item[Goal] The episode ends successfully. $\text{Reward}=1$.
\end{description}

Through trial-and-error, the agent gradually learns which actions are beneficial in a specific situation. Eventually, it will develop a policy that allows for navigating from \textit{Start} to \textit{Goal} without falling into \textit{Hole}s. By that, the policy encodes the MT sequence of interest, which can be materialized by selecting the actions (MTs) with the highest probability in each state.

The complexity of the problem prevents the agent from learning at a high pace. Typically, the agent struggles at the beginning of the learning process~\cite{sutton2018reinforcement}, especially when obstacles hinder exploration (e.g., the \textit{Hole} in \coord{1, 1}, and the \textit{Hole}s around \coord{2, 1}, forming a narrow passage).

\subsection*{Human guidance}

Human guidance can be of significant value to accelerate the agent's learning rate. For example, a human advisor might suggest the agent to avoid \coord{1, 1} as there is a hole (negative advice); or to visit \coord{2, 1} or \coord{1, 2}, because those tiles are part of every viable path between \textit{Start} and \textit{Goal} (positive advice).

The advisor might \textbf{not be completely certain} in the advice. For example, advisors might be more certain about the vicinity of their location (situational expertise), or might be experts in recognizing \textit{Hole}s but not \textit{Goal}s (domain-specific expertise).

\subsection*{Result of the learning process}

The result of the RL process is a policy that is able to assist the execution of MTs by providing priority information about which MT to execute in specific states, i.e., which direction to take while standing on specific frozen tiles. For example, the \texttt{moveRightRule} MT rule should be severely deprioritized when the vehicle is on tile \coord{0, 1} as it would lead to a termination by falling into a hole.

%% file: src/transformation-example.tex
\begin{lstlisting}[language={Xtend}, caption={Example \viatra{} MT rule moving the agent right}, label={lst:trafo-example}]
val moveRightRule = 
  createRule(agentOnFrozenTile)
 .name("MoveRightRule")
 .action[
    val newTile = tiles.findFirst[
      it.x==tile.x+1 && it.y==tile.y].head
    truck.tile = newTile].build
\end{lstlisting}

%% file: sections/approach.tex
\section{Overview of the approach}\label{sec:approach}

\subsection{Framework prototype}\label{sec:approach:framework}

To implement and evaluate our approach, we prototyped a framework. The goal of the framework is to support the registration of MTs or model manipulation primitives from which complex MT sequences are inferred via RL. The framework is available as an open-source software.\footnote{\url{https://github.com/ssm-lab/rl4mt}}

\begin{figure}[t]
    \centering
    \includegraphics[trim={0.75cm 0.5cm 0.75cm 0.5cm},clip,width=0.95\linewidth]{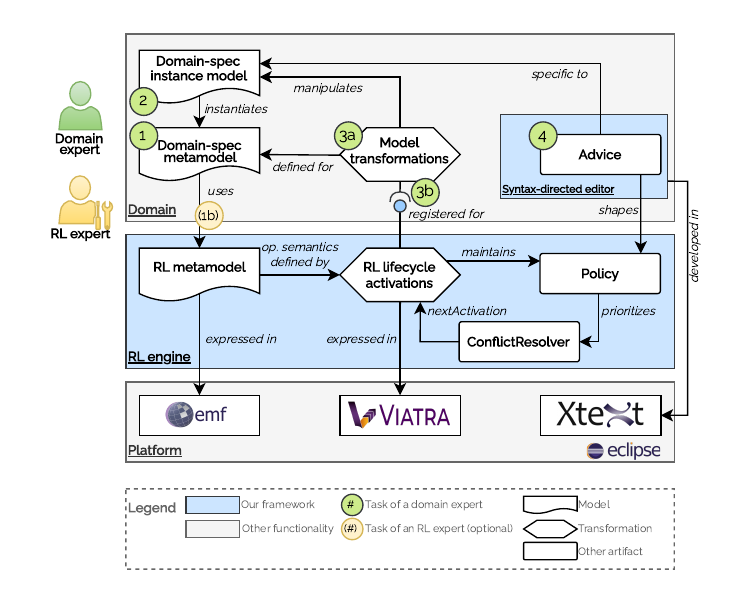}
    \caption{Architectural overview}
    \label{fig:architecture}
\end{figure}

\figref{fig:architecture} presents the architectural view of the framework (in blue), in the context of the typical domain concepts, activities, and users.
The core element of the framework is the \textit{RL Engine}. The engine is fully modeled: the \textit{RL metamodel} captures the domain-agnostic RL concepts (e.g., agent, policy, state) in the Eclipse Modeling Framework (EMF), and the operational semantics of RL (i.e., the underlying Markov Decision Process) is captured in MTs implemented on the \viatra{} platform. These MTs span the \textit{RL lifecycle activations} that drive the engine. (Described in detail in \secref{sec:approach:semantics}.) The \textit{Policy} is maintained by specific activations. To resolve ambiguous MT activations (i.e., when multiple MTs can fire), we implemented a \textit{ConflictResolver} that chooses the activation of the MT that is preferred in the Policy (i.e., has the highest probability in the state the agent exhibits).

\subsection{Domain-specific (meta)modeling}\label{sec:approach:modeling}

The framework allows for executing MTs as RL programs in an MDE process, with minimal development effort.

In step \colornumberdomainexpert{1}, a \textit{Domain-specific metamodel} is developed by the \textit{Domain expert} that reuses some key concepts of the \textit{RL metamodel}. Specifically, this step might be aided by an \textit{RL expert} in case the domain expert is not familiar with RL concepts. These cases are typical when a domain expert works with RL for the first time; the concepts to work with (e.g., terminating states and rewards) are trivial and could be circumvented by generative techniques.

\paragraph*{Mapping MTs to RL}
Related works have provided a clear mapping between MTs and RL~\cite{eisenberg2021towards,barriga2019personalized}. Here, we only recall that graph morphisms or MT left-hand sides (LHS) correspond to states in RL programs and model manipulations correspond to actions in RL. Reward is automatically calculated (e.g., number of violations in model repair~\cite{barriga2019personalized}), or assumes a reward structure (e.g., quality of solutions in model exploration~\cite{abdeen2014multi}).

In the running example, the domain expert may choose to refine the concept of the agent into the concept of the \textit{Truck}, and define the tiles on the lake as RL states, with frozen tiles being non-terminating states (the agent may continue the exploration) and holes and the goal being terminating states (the agent will stop exploring).

In Step \colornumberdomainexpert{2}, the domain expert instantiates the domain metamodel, e.g., as illustrated in \figref{fig:frozenlake}.

\subsection{Defining MTs}\label{sec:approach:mts}

In Step \colornumberdomainexpert{3a}, the domain expert defines the MTs for the domain model. In the running example, this means defining the four movements (left, down, right, up) for the autonomous vehicle. Once the MTs have been defined, they are registered in the RL engine in Step \colornumberdomainexpert{3b} to be executed as an RL program.

Typically, these MTs react to matched graph patterns expressed in terms of the domain-specific metamodel. This is why in the previous step, the domain expert (and in some cases, the RL expert) needed to ensure that the domain-specific metamodel uses some key concepts of the RL metamodel.
Alternatively, users of the framework may opt for defining plain model manipulations without an LHS, instead of complete MTs. Upon registering model manipulations, the framework packages them into MTs with a pattern in the RHS that describes non-terminating states in the RL agent's state space. This allows for the execution of model manipulations as long as the RL program does not terminate.

\subsection{Operational semantics}\label{sec:approach:semantics}

\begin{figure}[H]
    \centering
    \includegraphics[trim={0.75cm 0.35cm 0.66cm 0.5cm},clip,width=0.8\linewidth]{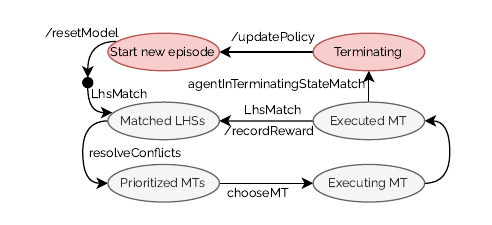}
    \caption{RL lifecycle for learning MT sequences}
    \label{fig:lifecycle}
\end{figure}

We developed our RL engine in a fully modeled fashion, in which the RL metamodel is operationalized by MTs. These MTs are internal to the RL engine and different from the registered domain-specific MTs. For example, updating the policy upon encountering a terminating state is formulated as an internal MT.
The activations of domain-specific and internal MTs span a lifecycle, shown in \figref{fig:lifecycle}.

The execution of the externally registered MTs starts by matching their LHS rules, resulting in some number of \textit{Matched LHSs}. Typically, multiple MTs' LHSs are matched, which requires resolving conflicts. In our implementation platform, \viatra{}, conflict resolution is achieved through a conflict resolver over a conflict set.\footnote{See \texttt{ConflictResolver.java} and \texttt{ConflictSet.java} on the API of \viatra{} in \url{https://github.com/eclipse-viatra/org.eclipse.viatra}.} We implement our conflict resolution logic in a way that MTs are dynamically prioritized by their probability in the policy. \lstref{lst:conflict-resolver} shows the relevant excerpt of the \texttt{PolicyBasedConflictSet} class in our implementation, which returns the next LHS activation.

\phantom{}

\input{src/conflict-resolver}

After resolving conflicts, the engine has a list of \textit{Prioritized MTs}. After choosing one of the prioritized MTs, the engine starts \textit{Executing the MT}. After execution, the result of the \textit{Executed MT} determines how the lifecycle proceeds.

If the agent is situated in a terminating state at this point, the lifecycle continues by \textit{Terminating} the episode, which includes updating the policy and \textit{Starting a new episode} by resetting the model (e.g., placing the agent in the initial state). This internal MT is shown in \lstref{lst:query} and \lstref{lst:trafo}.

\phantom{}

\input{src/iq-pattern-1}

\input{src/transformation-1}

If the agent is situated in a non-terminating state, the execution continues by recording the reward for the successful step and matching LHSs in the new state.

Through this lifecycle, the RL agent learns beneficial MT sequences based on the rewards for successful (non-terminating and rewarded) MT executions. The result of the RL process is a policy that can guide subsequent executions of the MTs to form the most efficient complex MT.

%% file: src/conflict-resolver.tex
\begin{lstlisting}[language={Xtend}, caption={Excerpt of the \textit{PolicyBasedConflictSet} class through which the policy-based conflict resolver prioritizes MTs}, label={lst:conflict-resolver}]
override getNextActivation() {
    val nextActivation =
          policyService.chooseAction(container)
    container.clear
    return nextActivation
}
\end{lstlisting}

%% file: src/iq-pattern-1.tex
\begin{lstlisting}[language={VQL}, caption={Domain-agnostic graph query matching morphisms in which the RL agent is in a terminating state}, label={lst:query}]
pattern agentInTerminatingState(
  agent: Agent,
  state: TerminatingState,
  environment: Environment){
    find agentInState(agent, state);
    Environment.agent(environment, agent);
}
\end{lstlisting}

%% file: src/transformation-1.tex
\begin{lstlisting}[language={Xtend}, caption={Domain-agnostic MT rule terminating the RL episode upon the agent encountering a terminating state}, label={lst:trafo}]
val agentIsTerminatedRule = 
  createRule(agentInTerminatingState)
 .name("AgentIsTerminatedRule")
 .action[agent.terminate].build
\end{lstlisting}

%% file: sections/guidance.tex
\section{Guiding RL agents by opinions}\label{sec:guidance}

In this section, we elaborate on Step \colornumberdomainexpert{4}.
Our previous work~\cite{dagenais2024opinion,dagenais2024towards} presents the general method in great detail. Here, we focus on adopting its essential elements required for this paper. We refer to \figref{fig:advice-approach} in the remainder of this section.

\subsection{Advisor input}\label{sec:approach-providing-advice}

\begin{figure}[t]
    \centering
    \includegraphics[trim={0.5cm 0.25cm 0.75cm 0.5cm},clip,width=\linewidth]{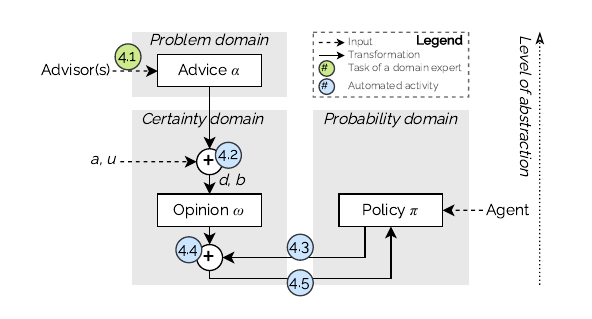}
    \caption{Method for incorporating advice into RL}
    \label{fig:advice-approach}
\end{figure}

First, the advisor provides advice to the agent. This step (shown in \colornumberdomainexpert{4.1} in \figref{fig:advice-approach}) is the only manual step in the method; the rest of the steps are automated in our technical framework.

The advisor expresses their subjective belief about a state being beneficial to the agent. We define advice as $\alpha: s \in S' \mapsto v$, i.e., a mapping from a state $s \in S'$ to a value $v$. Here, $S'\subseteq S$ is the subset of the problem space that the advisor can feasibly provide advice about, and $v$ is a value that describes the benefit of occupying or exhibiting that state. In the running example, a domain expert could formulate the advice ``\textit{Cell (1,1)} [a hole] \textit{is likely not beneficial}''. Here, $s$ refers to cell [1,1] in the grid world, and ``not beneficial'' is the value of occupying cell \cell{1}{1}. This advice is used to update the probability of choosing the action in adjacent states that leads to the advised cell, thereby shaping the policy.

Advice is provided through an appropriate domain-specific language (DSL)~\cite{schmidt2006model} and ideally, in a \textit{Syntax-directed editor}.
To demonstrate principles and conduct our experiments, we developed a DSL and an Xtext-based editor for the running example. Such editors can be developed or generated for the specific case at hand.
The DSL to provide advice about the grid world of the running example  is shown in \lstref{lst:gridworld-dsl}. An example set of advice expressed in the DSL about the running example (\figref{fig:frozenlake}) is shown in \lstref{lst:gridworld-dsl-1}.

\begin{figure}[h]
  \begin{lstlisting}[mathescape, language={gridworld-dsl}, caption={GridWorld grammar in EBNF}, label={lst:gridworld-dsl}]
<AdviceList> ::= <Advice>+
<Advice> ::= <AdviceLocation> `,` <AdviceValue>
<AdviceLocation> ::= `[`[0-9]+, [0-9]+`]`
<AdviceValue> ::= ^-?[0-2]
  \end{lstlisting}
\end{figure}

\begin{figure}[h]
  \begin{lstinputlisting}[mathescape, caption={Example set of opinions for the problem in \figref{fig:frozenlake}}, label={lst:gridworld-dsl-1}]{files/lake-1.txt}
  \end{lstinputlisting}
\end{figure}

The value of the advice represents the advisor's idea about how beneficial the given cell can be for the agent. In the example in \lstref{lst:gridworld-dsl-1}, the list of advice corresponds to the following cells of the running example (\figref{fig:frozenlake}): two holes on the grid, which are not beneficial (\parameter{1}{1}{-2}) and (\parameter{3}{1}{-2}); the goal, which is beneficial (\parameter{3}{3}{+2}); and a frozen tile that is not hazardous but still rather disadvantageous (\parameter{3}{0}{-1}) due to the holes in its vicinity.

\textbf{Uncertainty} in advice may arise due to various reasons. Typical to multi-view modeling settings, a subject-matter expert may be required to give advice about adjacent domains' concepts. For example, a mechanical engineer may need to express the quality of the electro-mechanical design.

\subsection{Compiling advice into opinion}\label{sec:advice-to-opinion}

Next, in step \colornumberauto{4.2}, advice is transformed into opinions by mapping to the parameters of \textbf{subjective logic}~\cite{josang2016subjective}---briefly introduced in \secref{sec:subjectiveLogic}.
This step is completely automated and is achieved in three sub-steps as follows.

\subsubsection{Calibrating base rate}
First, we set base rate $a_i$, which represents the prior probability of the agent taking an action in a state. For a set of actions \textit{A}, the base rate is derived as

{\small
\begin{align}
    a_i = \nicefrac{1}{|A|}. \label{eq:a}
\end{align}
}

In the running example, there are four actions the agent can choose in any state, thus $|A| = 4$ and the base rate $a_i= 0.25$. (Note that if the agent chooses an action that would lead it out of the grid world, it stays in place.) 

\subsubsection{Calibrating uncertainty}
Next, we calibrate uncertainty $u_i$. While uncertainty can be set manually by the advisor, it can prove difficult to quantify, and thus, we recommend deriving uncertainty using an automated method. 
For example, uncertainty may be derived from an appropriate distance metric, such that uncertainty increases with distance. To calibrate uncertainty using a distance measure, it is essential to choose (i) an appropriate distance metric and (ii) a discount function that successively discounts uncertainty as distance increases. 

In general the discount function is given as $\gamma: (u_{max}, \delta) \mapsto u_i \in (0, 1)$, where $\delta$ is a distance measured by a distance measure $\Delta$, and $u_{max}$ is the upper bound of uncertainty. From the identity of $u_i = 1 - (b_i+d_i)$ (\secref{sec:background}), it follows that $u_{max} = 1 - (b_i+d_i)$.
In the simplest case, a linear discount function can be chosen to calibrate uncertainty as follows.

{\small
\begin{align}
    u_i = \left(\nicefrac{\delta}{\delta_{max}}\right)\times u_{max} \label{eq:discount}
\end{align}
}

Here, $\nicefrac{\delta}{\delta_{max}}$ is the distance relative to the maximum distance, that maps onto the (0,1) domain.

In problems where uncertainty should be scaled only to a subset of the problem space, we use a threshold $0< \tau \in \Delta < 1$ to accelerate the discounting of certainty and reach $u_{max}$ in $\tau$. Then, \equref{eq:discount} is adopted as follows.

{\small
\begin{equation}\label{eq:h2-consistency}
  u_i =
  \begin{cases}
    \nicefrac{1}{\tau} \times \nicefrac{\delta}{\delta_{max}} \times u_{max} & \mbox{if } \delta \leq (\tau\times\delta_{max}); \\
    u_{max} & \mbox{if } \delta>(\tau\times\delta_{max}).
  \end{cases}
\end{equation}
}

\subsubsection{Computing disbelief (\textit{d}) and belief (\textit{b}) to form an opinion}\label{sec:approach-advice-compilation}
Finally, the belief $b_i$ and disbelief $d_i$ can be computed, and the parameters of the opinion $\omega_i$ can be set. As per the parameter constraints explained in \secref{sec:subjectiveLogic}, $u_i + b_i+d_i =1$, and $0 \le u_i, b_i, d_i\le 1$, so the remaining weight of $1-u_i$ is distributed between $b_i$ and $d_i$ based on the value of the advice $\alpha$. Given $n$, the length of the advice value scale, and the value of the advice $j$ (in ascending order of confidence from least to most), the calculation of $b_i$ and $d_i$ is given as 

{\small
\begin{align}
    b_i &= \left(\nicefrac{j-1}{n-1} \right)\times (1-u_i)~|~j \in \{1..n\} \label{eq:b} \\
    d_i &= (1-u_i) - b_i. \label{eq:d}
\end{align}
}

By using Equations \ref{eq:a}--\ref{eq:d}, the opinion $\omega_i$ is compiled as

{\small
\begin{align}
    \omega_i = \left(\frac{j-1}{n-1} \times (1-u_i), \frac{n-j}{n-1} \times (1-u_i), u_i, \frac{1}{|A|}\right). \label{eq:opinion}
\end{align}
}

The first parameter is the belief, $b_i$ as formulated in \equref{eq:b}, and the second parameter is the disbelief $d_i$ as formulated in \equref{eq:d}. The third parameter $u_i$ is the uncertainty calculated automatically by an appropriate metric. The final parameter $a_i$ is the base rate calculated by the dynamics of the problem space, as described in \equref{eq:a}.

Consider the advice from line 4 of \lstref{lst:gridworld-dsl-1}. Using an advice scale with length $n = 5$, the advisor indicates that the goal is beneficial $\alpha =$(\parameter{3}{3}{2}). Advice value $v=2$ corresponds with $j=5$. The agent can take one of the four actions in each cell in \figref{fig:frozenlake}, so $a_i = \nicefrac{1}{4} = 0.25$. This advisor has limited understanding of the environment, and is assigned uncertainty $u_i=0.5$. 
Using \equref{eq:opinion}, we calculate $\omega_i$ as

{\small
\begin{equation*}
    \begin{split}
        \omega_i &= \left(\frac{j-1}{n-1} \times (1-u_i), \frac{n-j}{n-1} \times (1-u_i), u_i, \frac{1}{|A|}\right)\\
        &= \left(\frac{5-1}{5-1} \times (1-0.5), \frac{5-5}{5-1} \times (1-0.5), 0.5, \frac{1}{|4|}\right)\\
        &= \left(0.5, 0, 0.5, 0.25\right).\\
    \end{split}
\end{equation*}
}

\subsection{Policy shaping via opinion fusion}\label{sec:policy-shaping}
Next, in the fully automated steps \colornumberauto{4.3} and \colornumberauto{4.4}, the agent's policy is shaped by the advisor's opinions as follows.

\subsubsection{Transforming Agent's Policy into certainty domain}
As described in \secref{sec:rl}, the agent maintains a policy  $\pi(a|s)$, which gives the conditional probability of choosing action $a \in A$ while in state $s \in S$. Since the entries are probabilities, $\forall s \in S, a \in A: 0 \leq \pi(a|s) \leq 1$. Before exploration, the probability of choosing any action from a state is equal, $\forall s \in S, a_i, a_j \in A, i \neq j: \pi(a_i|s) = \pi(a_j|s)$. 

To fuse opinions described in the certainty domain, with a policy described in the probability domain, we choose to transform the agent's policy to the certainty domain (step \colornumberauto{4.3}). Our choice is motivated by the well-defined and sound fusion operators that are readily available in the toolbox of subjective logic~\cite{josang2016subjective}.
This translation is achieved as follows:
$f_{P\rightarrow \Omega}(\pi): S \times A \rightarrow \Omega~|~P.$ As explained in \secref{sec:subjectiveLogic}, probability $p$ translates to opinion $\omega: p \mapsto (p, 1-p, 0, p).$

\subsubsection{Policy shaping by opinion fusion}
Next, in step \colornumberauto{4.4}, the advisor's opinions are fused with the agent's policy in the certainty domain. As described in \secref{sec:subjectiveLogic}, combined opinions can be fused into one joint opinion. By fusion, we are referring to a mapping $\Omega \times \Omega  \rightarrow \Omega$. For fusion, we require two elements: locating the policy entries that are affected by the advice (in the running example: the neighboring states of the advised state), and choosing an appropriate fusion operator from the toolbox of subjective logic.

\paragraph{Locating policy entries to be fused}
Since the advisor gives advice about the value of a particular state being beneficial, this advice will affect the agent's policy in states from which the advised state is reachable. The following definitions are required to formalize this step. 

\begin{definition}[Neighboring states]
    States $s_i, s_j \in S$ are said to be neighbors if there exists a state-action pair in the policy for which $\pi(a | s_i) \mapsto s_j$, i.e., choosing action $a \in A$ in state $s_i$ will transition the agent to state $s_j$. \label{def:neighbors}
\end{definition}

\begin{definition}[Neighborhood (of a state)]
    $N (s_i \in S) = \bigcup_{s_j \in S} \exists a \in A: \pi(a | s_i) \mapsto s_j$. \label{def:neighborhood}
\end{definition}
That is, neighborhood $N$ of a state is the set of all neighbors. In the running example, the state space is topological; thus, the neighborhood of a cell comprises the cells that can be reached from the given cell through one of the agent's actions.

Advice $\alpha$ about state $s_i$ is introduced to policy $\pi$ by $\forall a \in N(s_i).A : \omega(\alpha) \odot \omega(a)$. \label{theo:shaping}
That is, opinion $\omega(\alpha)$ formed from the advice is fused with every opinion formed from actions ($\omega(a)$) that lead from its neighboring states $N(s_i)$ to $s_i$.

\paragraph{Choosing a fusion operator}
To fuse opinions, \textcite{josang2016subjective} defines several fusion operators that are classified based on the situation. In this work, the agent and advisor formulate their opinions separately, and it is fair to assume they will not compromise. Thus, we use the Belief Constraint Fusion (BCF) operator, as it is most appropriate when compromise is not sought, and those giving opinions have made up their minds.
Joint opinion $\bcf{\omega} = (\bcf{b}, \bcf{d}, \bcf{u}, \bcf{a})$ under BCF is calculated by first finding the harmony and conflict of opinions: $\textit{Harmony} = b_1u_2 + b_2u_1 + b_1b_2$; $\textit{Conflict} = b_1 d_2 + b_2 d_1$.
Then, the components of the joint opinion are calculated as follows.

{\small
\begin{align}
    \bcf{b} &= \frac{\textit{Harmony}}{(1 - \textit{Conflict})}; &\bcf{u} &= \frac{u_1u_2}{(1 - \textit{Conflict})};\nonumber\\
    \bcf{d} &= 1 - (\bcf{b} + \bcf{u}); &\bcf{a} &= \frac{a_1(1-u_1) + a_2(1-u_2)}{2-u_1-u_2}.\nonumber
\end{align}
}

As described in \secref{sec:subjectiveLogic}, more than two opinions may be fused into a joint opinion. Due to this and the fact that fusion operators are commutative, our approach does not limit the number of advisors that may provide advice to the agent. 

\subsection{Transforming opinion-infused policies from the certainty domain to the probability domain}\label{sec:fused-to-policy}

Finally, in automated step \colornumberauto{4.5}, the agent's policy---now shaped by advisor opinions---is transformed from the certainty domain back to the probability domain.
First, as per \secref{sec:subjectiveLogic}, every opinion is transformed into a probability as $p: \omega \mapsto b + au$. Second, the policy needs to be normalized. State-specific policy entries form a probability space, and thus the invariant $\forall s \in S: \sum_{a \in A} p(a | s) = 1$ must hold. Normalization is defined as $\forall s \in S, a \in A: p(a | s) := p(a | s) \times \nicefrac{1}{\sum_{a \in A} p(a|s)}$.

The agent's policy now reflects the advice, and the agent may begin the exploration process. 

%% file: sections/evaluation.tex
\section{Evaluation}\label{sec:evaluation}

In this section, we evaluate our approach by answering the following research question: \textbf{how does human guidance affect the performance of RL in inferring MTs?}

\subsection{Evaluation setup}

We conduct a comparative study in which, first, we drive MTs by an unadvised RL agent, and then, we provide advice at various levels of uncertainty.
We use the $12\times12$ scaled-up, randomized version of the Frozen Lake environment (\figref{fig:frozenlake}). The map contains 20\% \textit{Hole} tiles, which is a standard ratio in OpenAI Gym's Frozen Lake implementation. The exact map is visualized in the replication package.
We assess the performance of RL agents by the cumulative reward they collect. We measure both the amount of the cumulative reward and the pace at which the RL agent collects it.
Our hypothesis is that human guidance has a measurable positive impact on both aspects, even at moderate levels of uncertainty.

\paragraph*{RL-specific parameters}
The RL-specific parameters of our experiments are shown in \tabref{tab:params-rl}. For the RL method, we choose discrete policy gradient~\cite{sutton2018reinforcement} as policy-based methods explicitly represent the policy and allow for the direct investigation of the effects of advice.
\textit{Learning rate $\alpha$} affects how much the agent updates its policy parameters during training. The closer it is to $0$, the smaller the policy parameter updates, and the closer it is to $1$, the larger the policy parameter updates. 
\textit{Discount factor $\gamma$} determines how much the agent values future rewards. The closer $\gamma$ is to $0$, the more the agent prioritizes immediate rewards, and the closer it is to $1$, the more the agent values long-term rewards.
Each experiment runs for a \textit{number of episodes}.
An episode starts by resetting the environment and placing the agent in the initial position. Within the same experiment, the policy is continuously maintained. An episode lasts until the agent either finds a terminating state (hole or goal) or reaches the number of \textit{maximum steps per episode}.
To mitigate threats to validity, we tuned these hyperparameters to favor the learning dynamics of the unadvised agent.

\input{tables/params-rl}

\paragraph*{Experiment parameters}
Experiment parameters are shown in \tabref{tab:params-experiment}.
We conduct 30 experiments for combinations of parameters and calculate their average to identify systemic tendencies. We compare the performance of different \textit{agent types}. The random agent chooses available actions randomly. The advised agent maintains and updates a policy. The unadvised agents also maintain and update a policy, and before exploration advice is incorporated into the policy. 

\input{tables/params-experiment}

We guide advised agents by different \textit{sources of advice} and \textit{advice quotas}. To form a ground truth, we first experiment with an idealized oracle with full information of the problem space, which provides advice based on pre-programmed rules. The oracle has two quotas; one where it provides advice about \textit{all} states (100\%) and another where it provides advice about the \textit{holes and goal} on the map (20\%). 

\input{tables/performance-synthetic}
\input{figures/figures-4x2}

In the \textit{single human mode}, one of the authors acted as the advisor with full information of the problem space. This mode has two advice quotas: one where advice is provided about 10\% of the state space, and the other where advice is provided about 5\% of the state space. For both the \textit{oracle} and \textit{single human} advisors, the \textit{degree of uncertainty} is modulated synthetically by setting $u = 0.0, 0.2, 0.4, 0.6, 0.8$. 

\paragraph*{Note about additional experiments}
The replication package contains additional data from experiments with two cooperating humans guiding the RL agent. Due to space restrictions, those results are not presented, but we note that \textbf{agents advised by cooperating humans outperform the unadvised agent} and \textbf{cooperating humans with lower quotas obtain similar performance to a single human advisor}.

\subsection{Results and key takeaways}

The results are shown in  \tabref{tab:cumulative-synthetic}
and are visualized in \figref{fig:cumulative-rewards-oracle-vs-human}.
In addition, we evaluate statistical significance between each advice mode at different levels of uncertainty. We use independent samples t-tests for the pairwise comparison. The replication package contains the complete list of t-test results.

We observe that \textbf{advised agents perform better than unadvised ones}, as in all but one case, advised agents accumulate more reward than the unadvised agent (mean reward 698.266). We also note that higher advice quotas tend to result in better performance. The main takeaway of this result is that \textbf{guidance, even if uncertain, improves the performance of the RL agent learning complex MTs}. 

We observe that the oracle with 100\% quota has the best overall performance, accumulating the most cumulative reward for all but one $u=0.0$, where the oracle with 20\% quota is slightly better. The \textbf{oracles with 100\% and 20\% quota, and the human advisor with 10\% have similar performance} for low uncertainty $u=0.0$ and $u=0.2$.
This is indicated by the cumulative reward and the similar slope in the log plots.
While the oracle with 100\% advice quota (\figref{fig:results-cumulativereward-synthetic-all}) significantly outperforms the human advisor with 10\% quota (\figref{fig:results-cumulativereward-human-10}) at low levels of uncertainty ($u \leq 0.4$), this difference is not significant at higher levels of uncertainty ($u \geq 0.6$). That is, \textbf{in uncertain situations, a human advisor, even with a lower advice quota, does not perform systematically worse than a hypothetical oracle}. This finding highlights the utility of human advice in uncertain situations.
This trend is even more pronounced when the oracle is given less advice quota (\figref{fig:results-cumulativereward-synthetic-holes}). With 20\% quota, the oracle does not significantly outperform a human advisor with 10\% quota when $u > 0.0$.

At low levels of uncertainty ($u \leq 0.4$) a human advisor with 10\% quota significantly outperforms the human advisor with 5\% quota. However, at higher levels of uncertainty ($u \geq 0.6$), the difference is not significant. These findings suggest that \textbf{different uncertainty levels require different advice strategies from human advisors}. Specifically, when advisors are certain, more advice is justified; however, when uncertain, advisors should be more frugal with their advice.

At low levels of uncertainty ($u \leq 0.4$), all advising modes (Fig. \ref{fig:results-cumulativereward-synthetic-all}--\ref{fig:results-cumulativereward-human-5}) significantly outperform the unadvised agent. However, as uncertainty increases ($u \geq 0.6$), the differences between advised and unadvised agents become statistically insignificant. These findings suggest that \textbf{guidance is beneficial with low-to-moderate uncertainty, but its utility is discounted as uncertainty increases}.

\begin{conclusionframe}{}
Our observations indicate that human guidance contributes to significant performance improvement in RL-based inference of complex MTs, even at moderate-to-high uncertainty. Human advisors, even with substantially lower advice quotas, can perform at the level of an oracle, and might even outperform it. This results in faster and more efficient inference of complex MTs.
\end{conclusionframe}

\subsection{Assumptions, limitations, and threats to validity}

\subsubsection{Assumptions and limitations}
We implicitly assumed that the reward structure could be effectively formulated. In some MDE applications, this is true. For example, \textcite{barriga2022parmorel} formulate reward by a user-preferred quality characteristic, such as maintainability and understandability.
Still, formulating reward might be challenging in open problems, such as design space exploration, in which reward often must be asserted to hypothetical, yet-to-encounter model configurations.

\subsubsection{Threats to validity}
A fortunately aligned reward structure may result in better, albeit non-systemic performance of the RL agent, threatening the \textit{internal validity} of our study. To mitigate this threat, (i) we have chosen the least supportive reward function in which the goal reward is minimal, there are no intermediate rewards, and terminating states do not provide negative feedback; and (ii) tuned the hyperparameters to favor the learning dynamics of the unadvised agent.
To assess the \textit{external validity} of our study, we remark that we only tested policy-based RL. While it is reasonable to expect the method to be applicable to other RL methods (e.g., value-based RL, a popular choice in MDE~\cite{barriga2022parmorel,eisenberg2021towards,sharbaf2022automatic}), our study does not provide evidence for this.
Similarly, a minor threat to external validity stems from the technological choices, especially the MT platform. We rely on the reactive capabilities of \viatra{}, and it is reasonable to expect that MT frameworks with support for similar MT execution semantics can adopt our approach.

%% file: tables/params-rl.tex
\begin{table}[t]
\centering
\caption{RL hyperparameters and settings}
\label{tab:params-rl}
\begin{tabular}{@{}ll@{}}
\toprule
\textbf{Parameter} & \multicolumn{1}{c}{\textbf{Value}} \\\midrule

RL method & Discrete policy gradient \\
Learning rate ($\alpha$) & 0.9\\
Discount factor ($\gamma$) & 1.0\\
Number of episodes & 10\,000 \\
Maximum steps per episode & 100 \\
State-action space & $12^2\times 4$\\
\bottomrule
\end{tabular}
\end{table}

%% file: tables/params-experiment.tex
\begin{table}[t]
\centering
\caption{Experiment parameters}
\label{tab:params-experiment}
\begin{tabular}{@{}ll@{}}
\toprule
\textbf{Parameter} & \multicolumn{1}{c}{\textbf{Value}} \\\midrule

Agent type  & \{Random, Unadvised, Advised\}\\
            

Source of advice    & \makecell[l]{\{Oracle, Single human\}}\\

Fusion operator & BCF\\

Advice quota -- Oracle  & \{100\%, 20\%\}\\

Advice quota -- Single human    & \{10\%, 5\%\}\\


Uncertainty (oracle and single human) & \{$0.2k~|~k \in {0..4}$\}\\



\bottomrule
\end{tabular}

\end{table}

%% file: tables/performance-synthetic.tex
\begin{table}[t]
\centering
\footnotesize
\captionsetup{width=\textwidth}
\caption{Cumulative rewards: Oracle and Single Human \\(Random = $0.033$, Unadvised = $698.266$.)}
\label{tab:cumulative-synthetic}
\begin{tabular}{@{}lrrrr@{}}
\cmidrule[\heavyrulewidth]{1-5}
 & \multicolumn{2}{c}{Oracle} & \multicolumn{2}{c}{Single human} \\
\cmidrule(r){2-3} \cmidrule(r){4-5}
u & \multicolumn{1}{c}{100\% (All)} & \multicolumn{1}{c}{20\% (H\&G)} & \multicolumn{1}{c}{10\%} & \multicolumn{1}{c}{5\%} \\
\cmidrule(r){1-1} \cmidrule(r){2-2} \cmidrule(r){3-3} \cmidrule(r){4-4} \cmidrule(r){5-5}
0.0 & 9\, 900.100 & 9\, 914.900 & 9\, 768.000 & 8\, 051.300\\
0.2 & 9\, 685.900 & 8\, 948.933 & 8\, 538.266 & 5\, 287.833\\
0.4 & 7\, 974.066 & 5\, 216.433 & 6\, 121.033 & 2\, 134.966\\
0.6 & 5\, 094.333 & 2\, 177.633 & 3\, 488.700 & 2\, 246.733\\
0.8 & 1\, 502.500 & 523.633  & 1\, 126.300 & 1\, 108.666\\
\cmidrule[\heavyrulewidth]{1-5}
\end{tabular}


\end{table}

%% file: figures/figures-4x2.tex
\begin{figure*}[t]
    \centering

    \caption*{Linear scale}

    
    \begin{subfigure}{0.235\linewidth}
        \includegraphics[width=\linewidth]{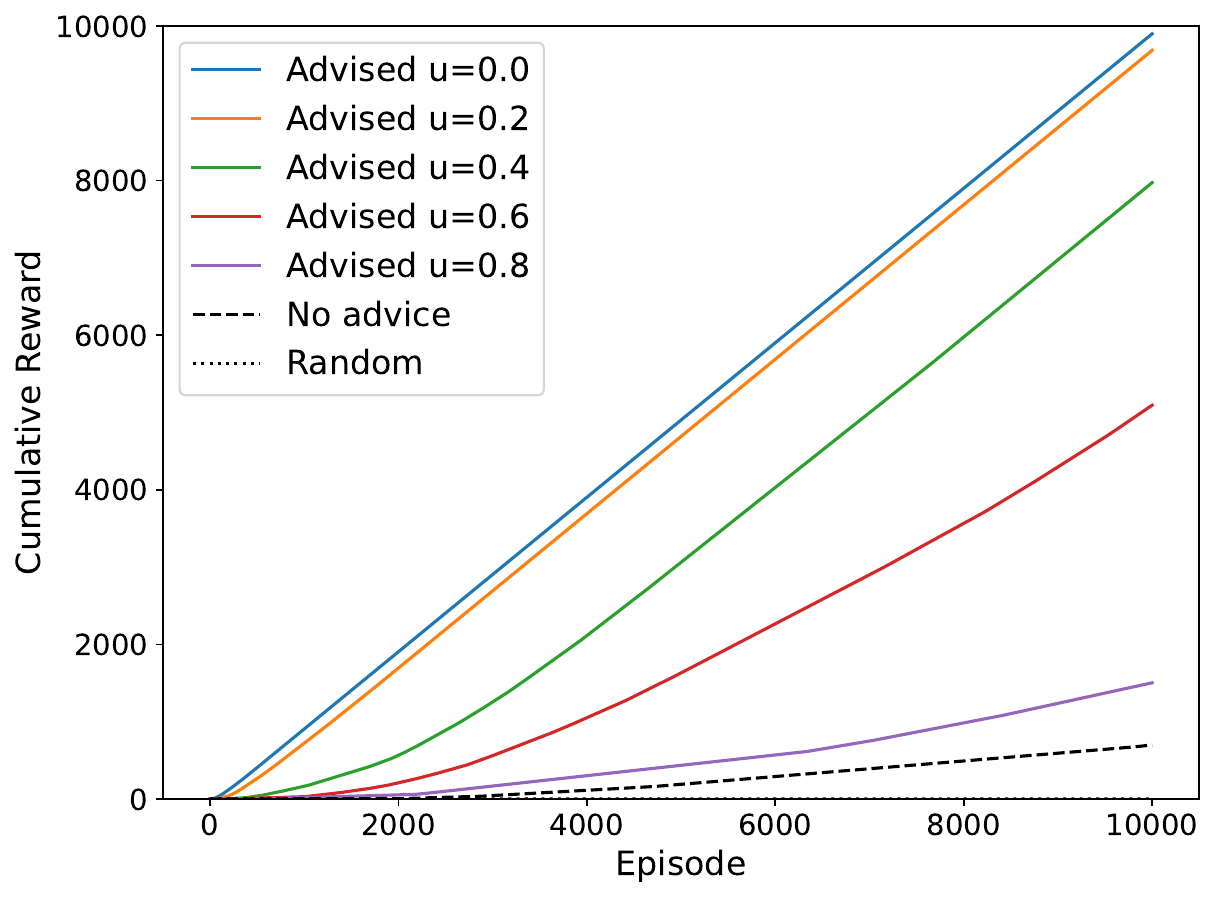}
        \caption{Oracle 100\%}
        \label{fig:results-cumulativereward-synthetic-all}
    \end{subfigure}
    \hfil
    \begin{subfigure}{0.235\linewidth}
        \includegraphics[width=\linewidth]{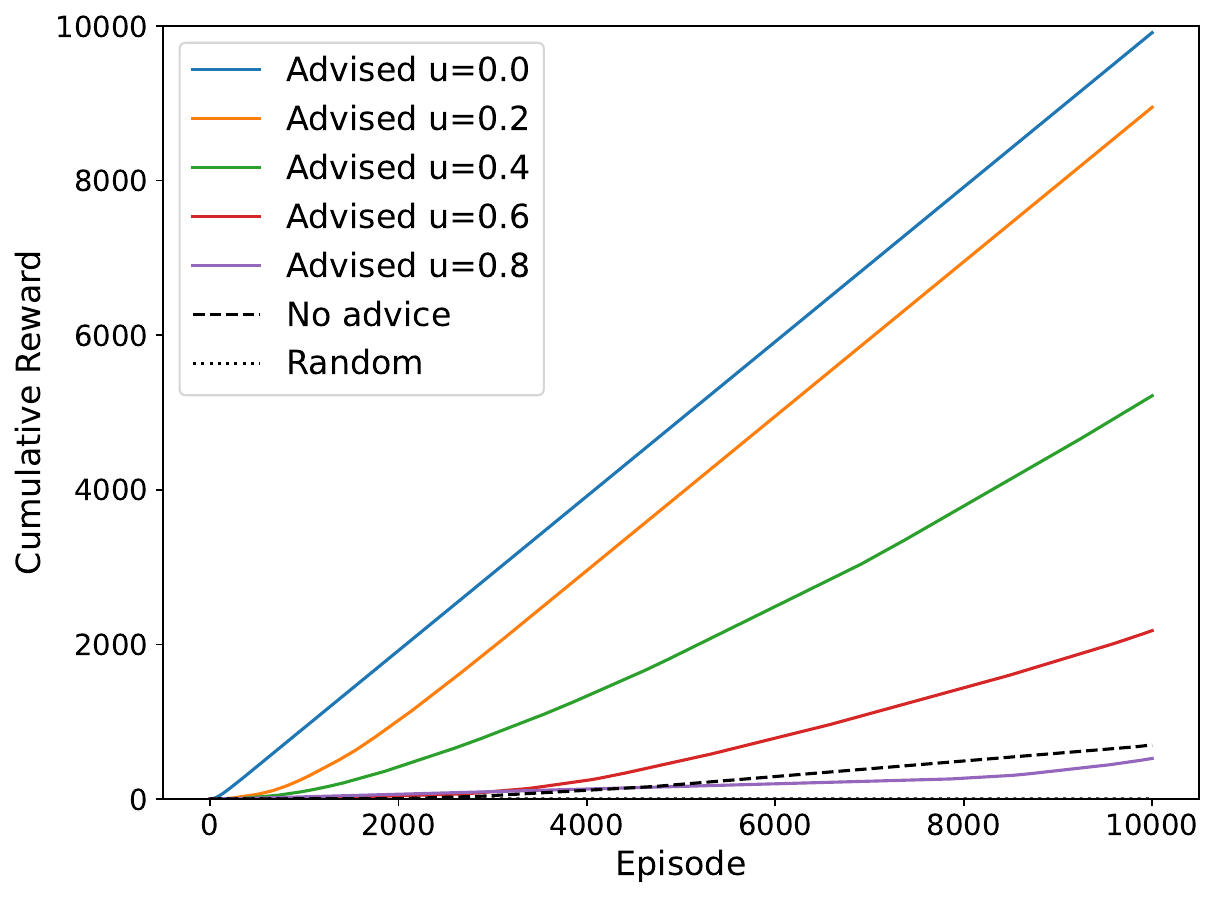}
        \caption{Oracle 20\%}
        \label{fig:results-cumulativereward-synthetic-holes}
    \end{subfigure}
    \hfil
    \begin{subfigure}{0.235\linewidth}
        \includegraphics[width=\linewidth]{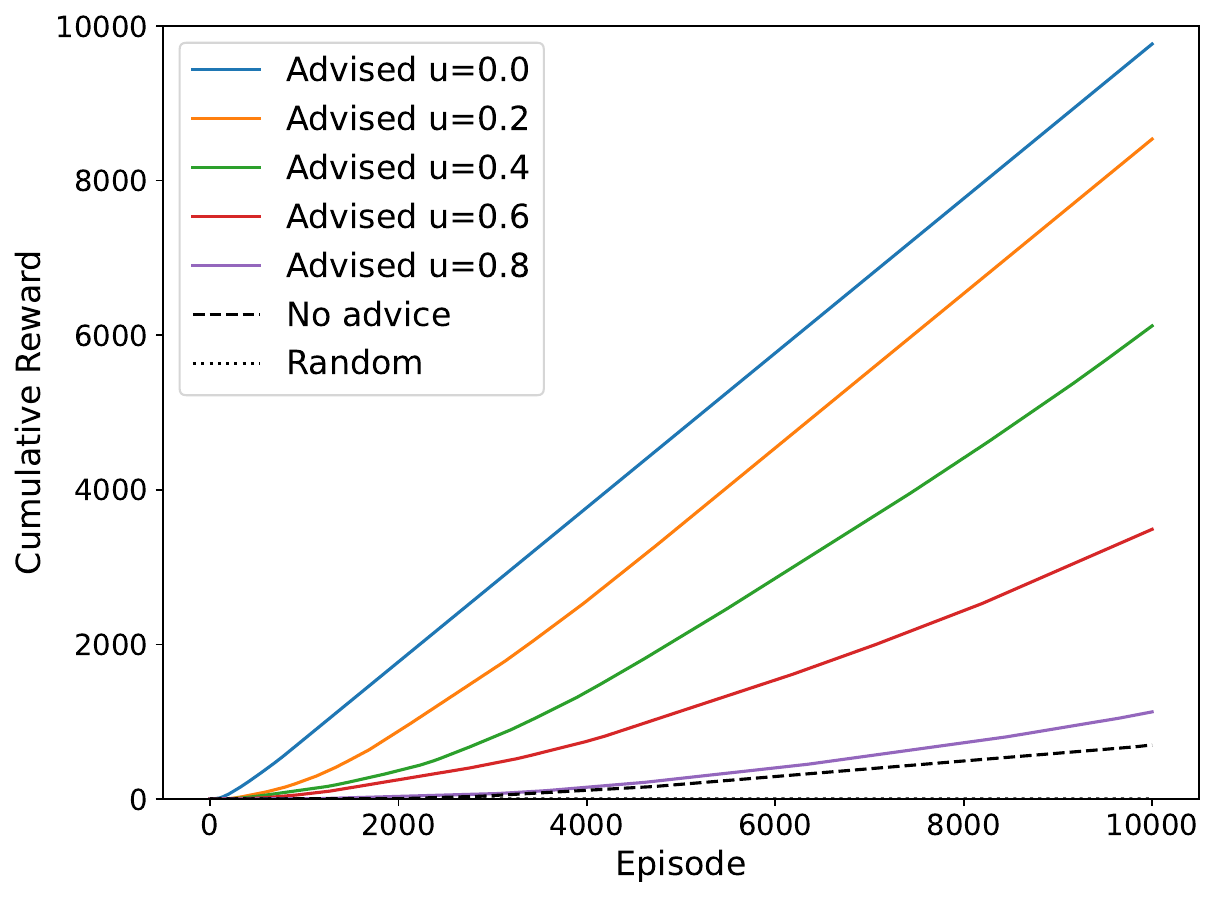}
        \caption{Human 10\%}
        \label{fig:results-cumulativereward-human-10}
    \end{subfigure}
    \hfil
    \begin{subfigure}{0.235\linewidth}
        \includegraphics[width=\linewidth]{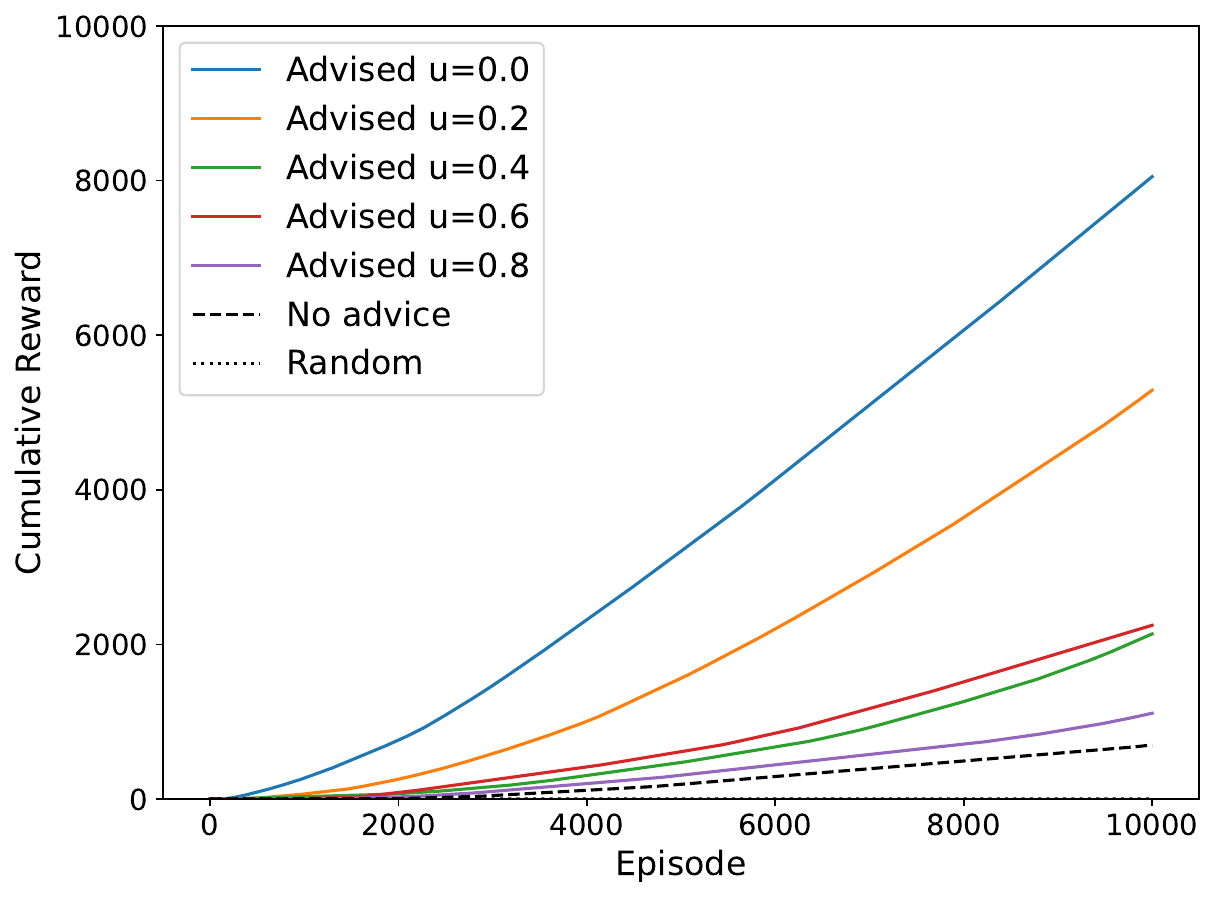}
        \caption{Human 5\%}
        \label{fig:results-cumulativereward-human-5}
    \end{subfigure}


    \caption*{Log scale}


    \begin{subfigure}{0.235\linewidth}
        \includegraphics[width=\linewidth]{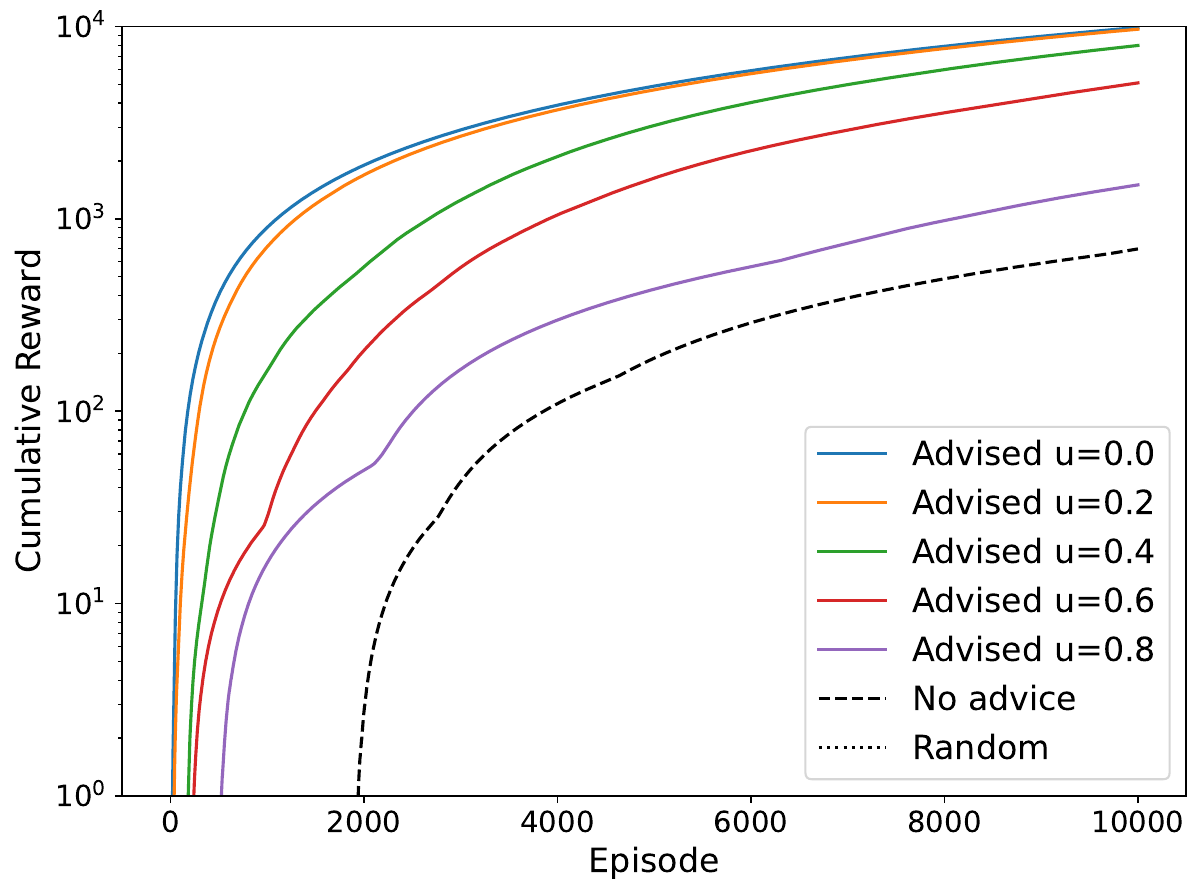}
        \caption{Oracle 100\%}
    \end{subfigure}
    \hfil
    \begin{subfigure}{0.235\linewidth}
        \includegraphics[width=\linewidth]{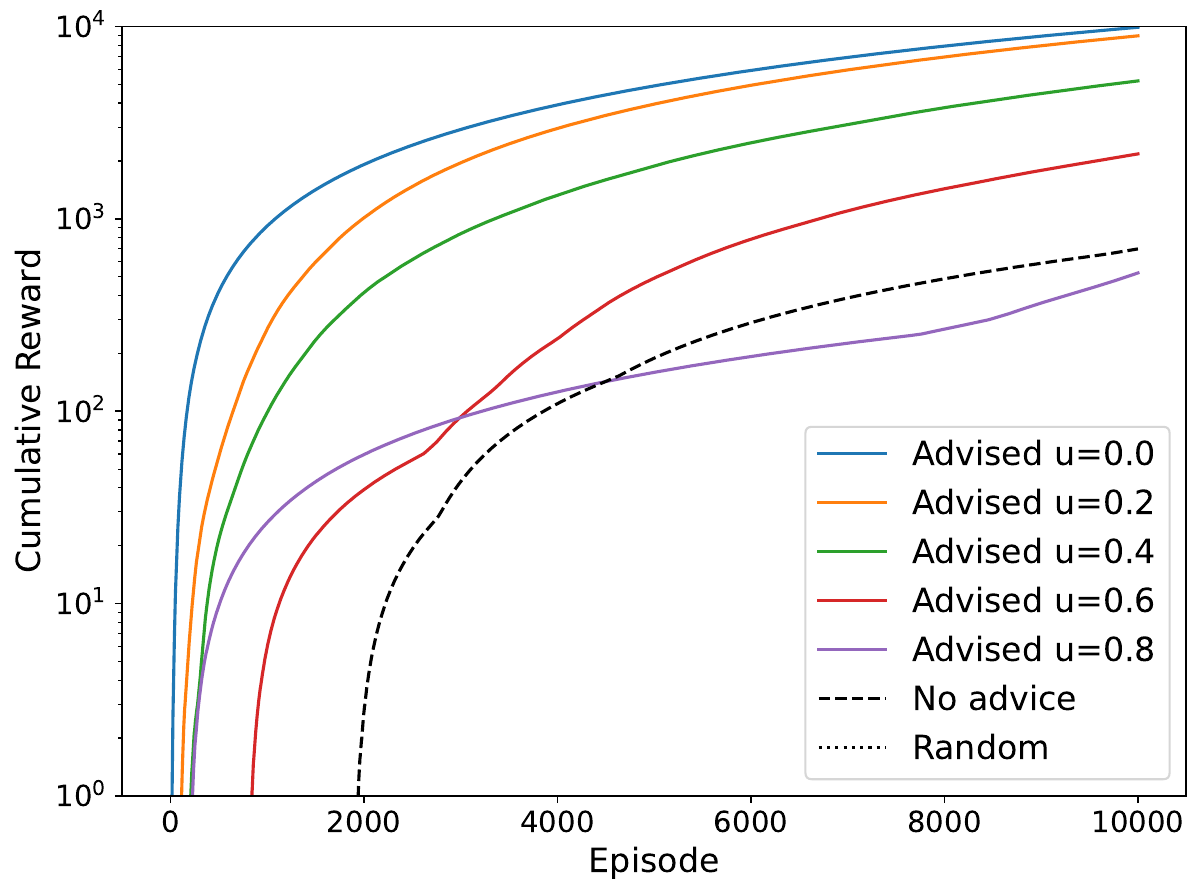}
        \caption{Oracle 20\%}
    \end{subfigure}
    \hfil
    \begin{subfigure}{0.235\linewidth}
        \includegraphics[width=\linewidth]{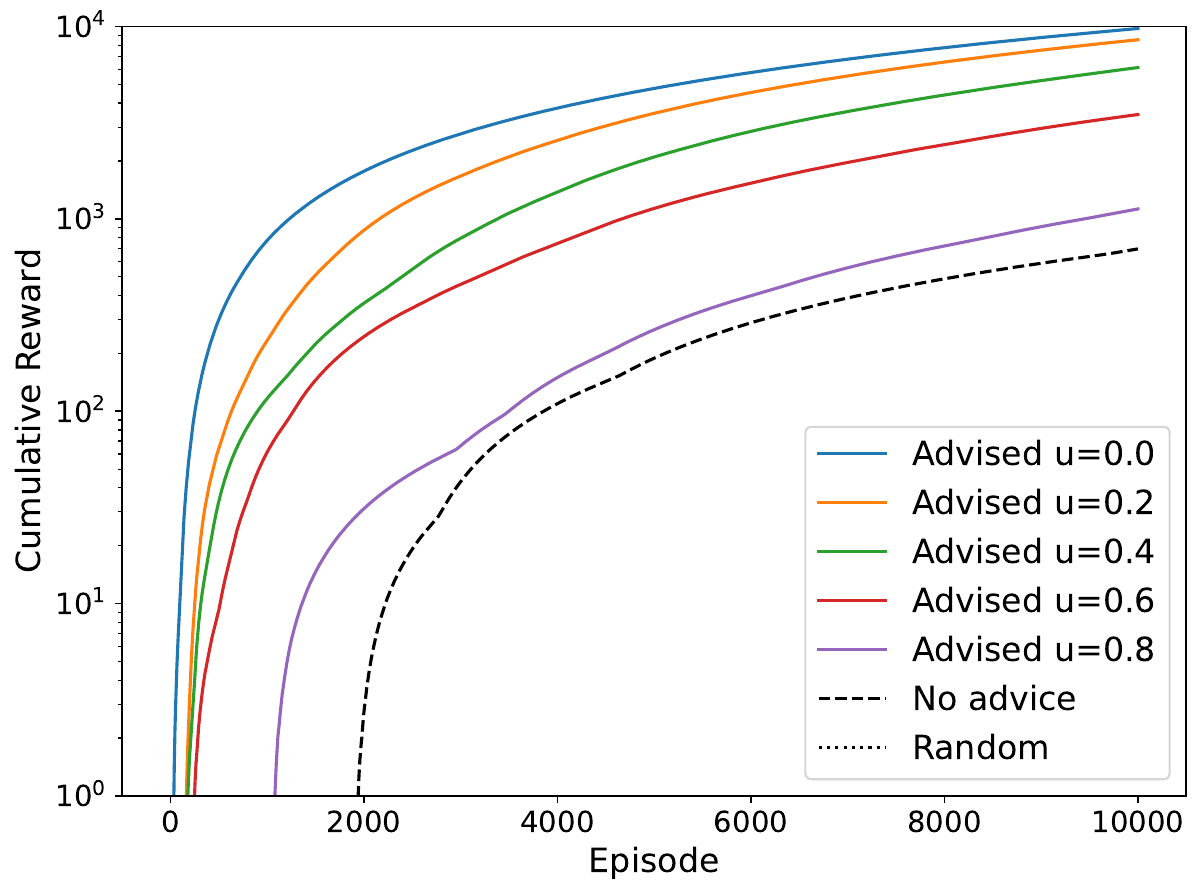}
        \caption{Human 10\%}
    \end{subfigure}
    \hfil
    \begin{subfigure}{0.235\linewidth}
        \includegraphics[width=\linewidth]{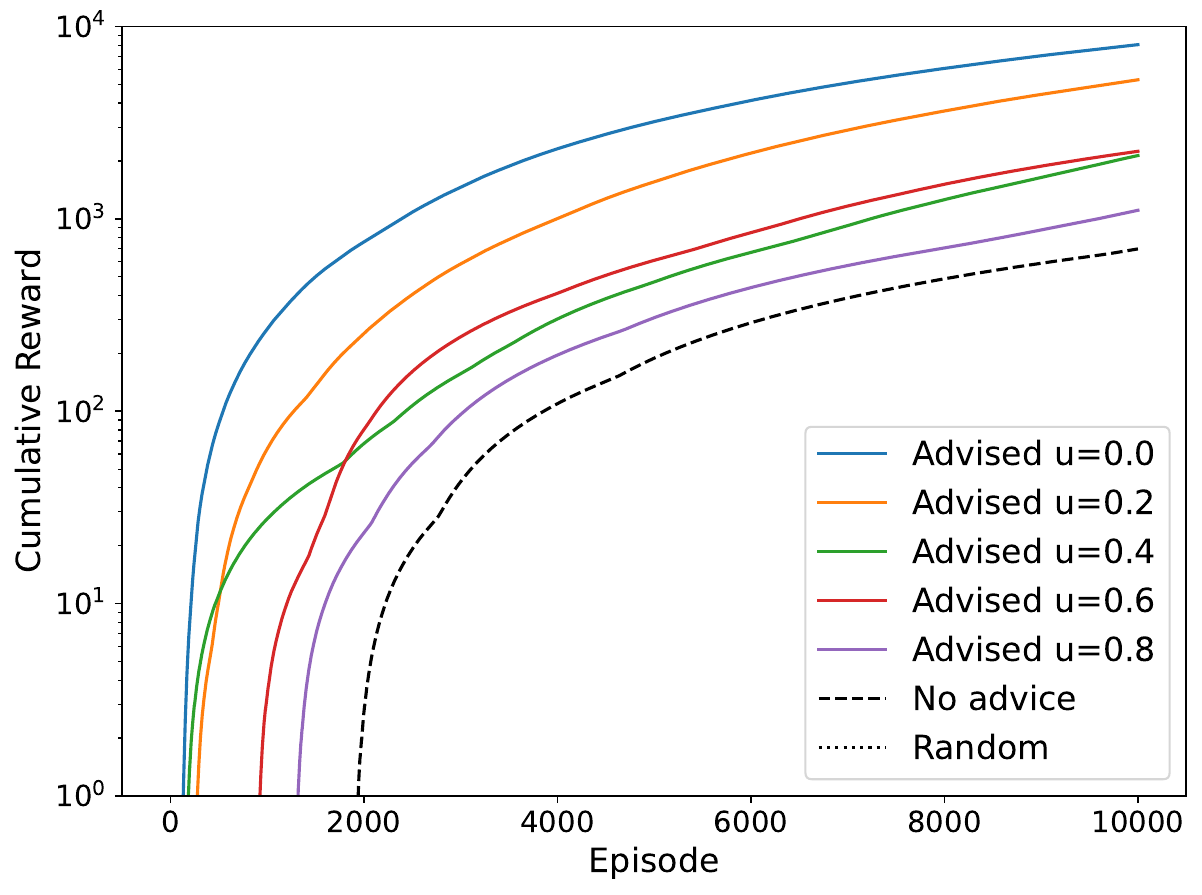}
        \caption{Human 5\%}
    \end{subfigure}

    \caption{Performance comparison of synthetic oracles and real human advisors in terms of cumulative reward}
    \label{fig:cumulative-rewards-oracle-vs-human}


\end{figure*}

%% file: sections/discussion.tex
\section{Discussion}\label{sec:discussion}

We now discuss the implications of our work and outline areas of interest for MDE researchers.

\subsection{Towards RL-enabled MT platforms}

The result of the RL process is a policy that encodes beneficial MT executions in specific graph morphisms. To utilize this information, we foresee two main directions.

\subsubsection{Runtime policy exploitation in MT chains} The RL engine can be used as a facility to resolve MT activations based on the policy. The policy asserts higher probability to specific MTs in a model state due to the higher positive reward the MT yielded in previous situations of exhibiting the model state. Such scenarios can be supported through appropriate conflict resolves that factor in the policy information, as shown in \figref{fig:architecture} and \figref{fig:lifecycle}. The benefit of this approach is that the RL engine can continuously maintain the complex MT and guide the execution of complex MTs in a non-intrusive way.

\subsubsection{Generating standalone MT chains} Alternatively, standalone MT chains can be generated to be used independently from the RL engine. In such scenarios, the RL engine is used only for training purposes and does not maintain the MT chain.

We anticipate MT platform developers to adopt techniques such as ours and promote RL-enabled MTs to a first-class citizens in their platforms. While the former approach leverages our approach to a greater extent, the latter one is less intrusive and requires a smaller leap for the user base of MT platforms.

\subsection{Typical applications}

To better contextualize our approach within MDE, we explain how some characteristic MDE problems map onto RL and can make use of our approach.

\subsubsection{Model synchronization and inconsistency management}
These problems are related and both often require identifying complex MTs~\cite{mens2007incremental,elhamlaoui2018mde,david2016engineering}. In these settings, the RL \textbf{state} encapsulates morphisms over the union of the source and target models, and the \textbf{reward} is typically implied by the negative distance between the source and target models. The lower the distance, the better the solution. Termination semantics are asserted to states that minimize the distance or, equivalently, maximize the reward. Distance can be measured at the syntactic level (e.g., by counting inconsistent model elements)~\cite{marchezan2024tool} or at the semantic level (e.g., by counting inconsistent properties that are not being jointly satisfied by the two models)~\cite{david2023model}.

\paragraph*{The role of our approach} \textbf{Actions} are typically expressed as domain-specific MTs, but are limited in complexity. Our approach allows for inferring complex MTs that allow for more sophisticated synchronization semantics between models.
\textbf{Human guidance} is of particular value in these problems as multi-domain settings challenge the understanding of how exactly consistency should be restored~\cite{jongeling2023uncertainty,dagenais2024driving}.

\subsubsection{Model repair and refactoring}

This class of problems is similar to model synchronization; however, there are some key differences. RL \textbf{state} in this case encapsulates only a single model's morphisms, and \textbf{reward} cannot be automatically calculated due to the lack of a suitable oracle. That is, it cannot be structurally defined when a refactoring is successful or sufficient. Instead, the reward structure must be defined by the domain expert. Typically, qualitative metrics are used for this purpose, either defined at the general syntactic level (e.g., number of dangling edges or isolated components)~\cite{taentzer2017change-preserving} or at the domain-specific syntactic level~\cite{syriani2019domain-specific}. Occasionally, personalized metrics are considered~\cite{barriga2022parmorel}.

\paragraph*{The role of our approach} The challenge is similar to the previous class of problems: \textbf{actions} are usually expressed as domain-specific MTs with limited complexity. Here again, an approach such as ours would allow for inferring complex MTs that, in turn, enable more complex refactorings and model repair actions. \textbf{Human guidance} helps prevent models from growing out of size under the premise of repair.

\subsubsection{Design space exploration}

Design space exploration (DSE) is the process of identifying favorable candidate designs under various constraints and soft objectives~\cite{herzig2017model-transformation-based}. Starting from an initial design, a DSE process gradually applies MTs to uncover alternative designs. In such scenarios, the \textbf{state} encapsulates a specific graph morphism in a model instance. \textbf{Actions} are typically defined in domain-specific terms~\cite{hegedus2015model} rather than domain-agnostic CRUD primitives. The \textbf{reward} structure is given through constraints and soft objectives.

\paragraph{The role of our approach} Optimal solutions are often found deep in the design space and are computationally demanding to identify. Our approach can aid the computational load of DSE by learning complex MTs that have led to beneficial design alternatives. These complex MTs, then, can be used as fixed points in the design space from which subsequent DSE executions may continue deeper explorations. \textbf{Human guidance} helps by providing known favorable design decisions and by that, drive the RL agent to favorable states and accelerate the inference of complex MTs. The challenge of uncertainty in DSE has been recognized before~\cite{foldiak2024probabilistic}. Our approach allows for the explicit modeling and management of uncertainty that stems from human guidance.

\subsection{Research opportunities for the MDE community}

There are some key challenges to be addressed in order to materialize the opportunities our approach creates. In the following, we elaborate on the ones that align particularly well with the expertise and knowledge of the MDE community.

\subsubsection{Domain-specific languages for guidance}

A key advantage of opinion-based guidance is the rapid pace at which opinions emerge. To leverage this advantage to its fullest, intuitive languages are needed to express opinions. Domain-specific languages (DLSs)~\cite{schmidt2006model} have been widely used to capture expert opinions at high levels of abstraction, narrowing the gap between the modeling language and human cognition. However, DSLs typically operate with reduced syntax to remain truly specific to the problem domain, and therefore, \textbf{automated methods for DSL engineering} are in high demand~\cite{bucchiarone2020grand}. This is true for guided RL as well. We envision language workbenches that allow for the automated construction of DSLs through which advice can be provided by domain concepts. In the running example (\secref{sec:example}), one might want to use descriptive quantifiers, such as ``hazardous'' or ``beneficial'' instead of relying on an integer scale.
Unfortunately, there is little research on \textbf{domain-specific languages for RL}, and existing languages mostly focus on programming RL procedures~\cite{molderez2019marlon} rather than expressing opinions at high cognitive levels. We see plenty of room for research targeting domain-specific modeling languages for RL, inspired by the advancements in the model-driven engineering community~\cite{kulagin2022ontology,wu2005automated}.

\subsubsection{(Domain-specific) Tool support}

Our approach can be aided by domain-specific tool support at key points. For example, the domain expert needs a \textbf{domain-specific editor} with an intuitive user interface to efficiently provide advice. We implemented the prototype of such an editor on top of Eclipse and Xtext, as discussed in \secref{sec:approach-providing-advice}. Instead of case-by-case tool development,
syntax-directed editors could be generated from high-level specifications of the RL problem and the application domain~\cite{syriani2021generation}, and equipped with automatically generated specialized concrete syntax~\cite{benchaaben2024intelligent}. Adaptive modeling languages~\cite{delara2025adaptive} offer an apt solution through a product line-inspired approach to customizing DSLs to different problems.

Additional tooling considerations relate to \textbf{model management} and RL execution. In our work, we executed the RL agent's learning process on an instance model that should likely be prevented from automated complex MTs until the RL agent is trained. Instead, an improved version of our tool could create replicas of the real model and treat those as sandbox models for the agent to train. The result of the training, i.e., the policy, can be applied on the real model. Such utility functionality will pave the way for adoption.

\subsubsection{Interactive advice}
In our experiments, we assumed that advice is provided once, before the agent begins exploring the environment. However, given adequate interaction protocols, our approach can accommodate advice at any point during the RL process. We see opportunities in developing more \textbf{interactive advice protocols} through advanced human-computer interactions (HCI)~\cite{shajari2025bridging}.
MDE can be of high value in such situations, as it can support \textbf{domain-specific interfaces} that integrate user models in the system design. This would allow for interactive advising systems that are tailored to the user's level of expertise about the problem space~\cite{abrahao2017user}.
Blended modeling---the activity of interacting seamlessly with a single model (i.e., abstract syntax) through multiple notations (i.e., concrete syntaxes), allowing a certain degree of temporary inconsistencies~\cite{david2023blended}---can be of high utility in facilitating seamless advising workflows in multi-view settings.

Finally, interaction protocols amenable to \textbf{verification and validation} can be generated, e.g., from state chart models~\cite{harel1987statecharts} or sequence diagrams~\cite{kundu2013automatic}.

%% file: sections/conclusion.tex
\section{Conclusion}\label{sec:conclusion}

We presented a method for engineering complex model transformation (MT) sequences by reinforcement learning (RL), guided by potentially uncertain human advice. Our results show that human advice, even if uncertain, can be a major contributor to more efficient MT development. Our results open the door for the RL-based human-in-the-loop treatment of key MDE problems, e.g., model synchronization and design-space exploration.
We envision MT platform developers adopting techniques such as ours to promote RL-backed MTs to first-class citizens in MT platforms. This will enable efficient MT development, execution, and maintenance.

Future work will focus on evaluating our approach for different flavors of RL and developing domain-specific support
for RL-driven human-in-the-loop MT engineering.

%% file: sections/acknowledgement.tex
\section*{Acknowledgement}

We acknowledge the support of the Natural Sciences and Engineering Research Council of Canada (NSERC), DGECR-2024-00293 (End-to-end Sustainable Systems Engineering).